\documentclass[twocolumn,prl,aps,showpacs,superscriptaddress]{revtex4}
\usepackage{epsfig}
\usepackage{graphicx}

\usepackage{epstopdf}
\begin{document}

\title{Acoustic Noise of MRI Scans of the Internal Auditory Canal and Potential for Intracochlear Physiological Changes}

\author{M.~A.~Busada}
\author{C.~L.~Eshleman}
\author{G.~Ibrahim}
\author{J.~H.~Huckans} \email{jhuckans@bloomu.edu}

\affiliation{College of Science of Technology, Bloomsburg University of Pennsylvania, Bloomsburg, Pennsylvania, 17815, USA}

\date{\today}

\begin{abstract}
Magnetic resonance imaging (MRI) is a widely used medical imaging technique to assess the health of the auditory (vestibulocochlear) nerve (VCN).  A well known problem with MRI machines is that the acoustic noise they generate during a scan can cause auditory temporary threshold shifts (TTS) in humans\cite{Randomskij,Counter,Ulmer,McJury}.  In addition, studies have shown that excessive noise in general can cause rapid physiological changes of constituents of the VCN within the cochlea\cite{Wang}.  Here, we report in-situ measurements of the acoustic noise from a 1.5 Tesla MRI machine (GE Signa) during scans specific to VCN assessment\cite{Panas}.  The measured average and maximum noise levels corroborate earlier investigations where TTS occurred.  We briefly discuss the potential for physiological changes to the intracochlear branches of the VCN as well as iatrogenic misdiagnoses of intralabyrinthine and intracochlear schwannomas due to hypertrophe of the VCN within the cochlea during MRI assessment.

\end{abstract}


\maketitle

An acoustic neuroma is a type of abnormal hyperplasia involving the Schwann cells surrounding the vestibulocochlear nerve (VCN). Intralabyrinthine schwannomas (ILSs) are acoustic neuromas of the intralabyrinthine branches of the VCN.  Unlike most schwannomas, they are not located in the internal auditory canal (IAC)\cite{Tieleman,Neff}.  Intracochlear schwannomas (ICSs) grow on the intracochlear branches of the VCN.  Magnetic resonance imaging (MRI) can now routinely detect acoustic neuromas as small as 1.0 mm in size\cite{Friedman}.  With the advent of gadolinium (Gd)-enhanced MRI, more ILSs and ICSs are being detected during screening for hearing loss, vertigo or tinnitus\cite{Brogan}.  Still, little is known about the prevalence of ILSs and ICSs; less is known about their growth patterns \cite{Tieleman}. Further improvements in MRI resolution will make more precise localizations of small ILSs and ICSs possible.  Increased imaging ability demands assessing the possible physiological effects of MRI machine acoustic noise on the inner ear.

Here, we report detailed measurements and analysis of the intra-bore acoustic noise generated by a 1.5 Tesla GE Signa MRI machine used for standard auditory nerve assessment at Bloomsburg Hospital in Bloomsburg, Pennsylvania.  These measurements are compared with prior acoustical studies\cite{Randomskij,Counter,Ulmer,McJury} which documented auditory threshold shifts in humans as a function of noise levels. Comparisons are also made to an earlier histological study\cite{Wang} on the physiological effects of intense noise on murine inner ears.

\begin{figure}[htbp]
   \centering
   \includegraphics [width=8.5cm,height=8cm]{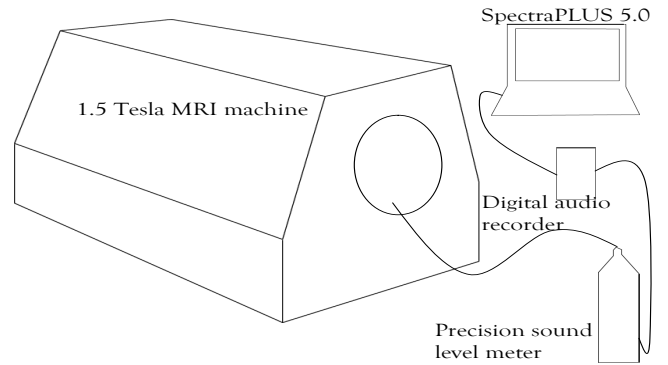}
   \caption{Acoustical noise generated by 1.5 Tesla MRI machine was channeled via a 5 m long flexible tube to a precision sound level meter whose AC line output was recorded on a digital audio recorder for subsequent time and frequency analysis (SpectraPLUS 5.0).}
   \label{Figure:figex1}
\end{figure}

To make in-situ acoustical measurements of the 1.5 Tesla GE Signa MRI machine during its scan, it was not possible to locate the metallic measurement apparatus directly within its bore because of the intense magnetic field.  Therefore, a technique was developed to channel the sound within the MRI bore down a tube to the remote measuring equipment.  As shown in Fig.~\ref{Figure:figex1}, the open end of an approximately 5 m long flexible tube (inner diameter $\frac{1}{2}$ in.) was positioned at the side of a phantom head within the MRI bore.  The sound then traveled down the tubing where, at the other end, the tube was connected to a $\frac{1}{2}$ in. ANSI S1.15 Type 1 precision microphone\cite{aco} and ANSI S1.4 Type 1 impulse precision sound level meter\cite{bruelslm} with its AC line output recorded on a digital audio recorder\cite{zoom}.

\begin{figure}[htbp]
   \includegraphics[width=8.5cm, height=8cm]{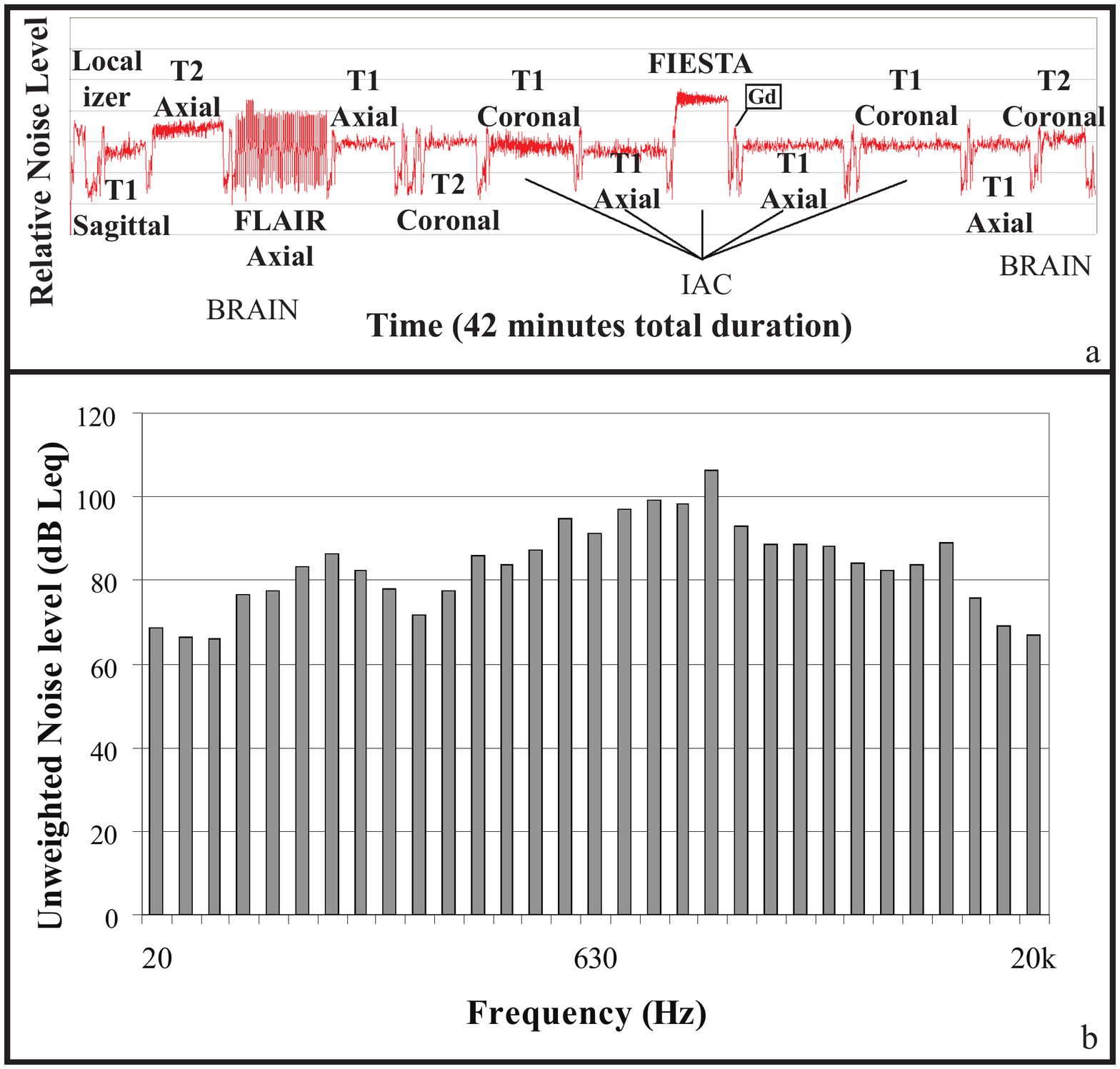}
   \caption{a) A representative MRI acoustical noise time series measurement with particular assessment sequences labeled. b) The 42-minute Leq (non A-weighted) spectrum at the location of the phantom ear (i.e. with the application of the flexible tubing transfer function).}
   \label{Figure:figex2}
\end{figure}

The frequency network and detector response of the sound level meter were set to linear and fast, respectively.  The measurement system was field-calibrated immediately prior to each measurement using a NIST-traceable acoustical calibrator\cite{bruelcal}.  The digital files from the recorder were post-processed using audio spectrum analysis software\cite{spectraplus}.  The acoustical transfer function of the tubing was separately measured using a pink noise source and then applied to the MRI measurements.  A representative MRI acoustical noise time series measurement is shown in Fig.~\ref{Figure:figex2}a.

The thirteen sequences comprising the complete auditory assessment protocol lasted a total duration of 42 minutes.  This resulted in an unprotected equivalent noise level of 109 dBA $\rm{L_{eq}}$ after applying the transfer function of the flexible tubing and A-weighting.  The measured $\rm{L_{10}}$, $\rm{L_{50}}$ and $\rm{L_{90}}$ exceedance levels were 111 dBA, 106 dBA and 100 dBA, respectively.  The maximum noise level was 124 dBA. Fig.~\ref{Figure:figex2}b presents the 42-minute $\rm{L_{eq}}$ spectrum (non A-weighted) calculated at the phantom ear (i.e. with the application of the flexible tubing transfer function).  At Bloomsburg Hospital, FIESTA (fast imaging employing steady state acquisition), also known as CISS (constructive interference steady state), is one of the most important MRI sequences to identify an acoustic neuroma.  Note in Fig.~\ref{Figure:figex2}a that it is also one of the loudest components of the scan.  The complete auditory nerve assessment protocol normally performed by Bloomsburg Hospital is summarized in Table I.

Comparisons can be made between our measured noise levels and established standards for permissible noise exposure such as those promulgated by the Occupational Safety and Health Administration (OSHA) \cite{OSHA}.  The purpose of the OSHA noise exposure standards is to protect individuals from habitual exposure to excessive noise levels that could cause hearing damage.  On an eight-hour OSHA basis, we measured an unmitigated noise dosage of 92 dBA which exceeds the 90 dBA OSHA noise exposure criterion.

\begin{table}[b!]
\label{Experimental Data table}
\vspace{10pt}
\begin{tabular}{c|c|c|c|c|c}

Sequence               &  Echo    & Repetition        & Slice        & Duration    & $\rm{L_{eq}}$    \\
name                   &  time, TE& time, TR          & thickness    & (min:sec)   & (dB)    \\
                       &  (ms)    & (ms)              & (mm)         &             &        \\
\hline
Brain target\\
\hline
3D localizer           &           &                  &              & 0:26        & 100.8    \\
T1 sagittal            &  9        & 350              & 5            & 1:35        & 96.7    \\
T2 axial               & 85        & 4325             & 5            & 2:39        & 100.6    \\
FLAIR axial            & 120       & 10000            & 5            & 3:20        & 97.7    \\
T1 axial               & 17        & 500              & 5            & 2:12        & 97.3    \\
T2 cor - fat             & 85        & 4400             & 5            & 2:07        & 98.1    \\
\hline
IAC target\\
\hline
T1 coronal             & 14        & 450              & 3            & 3:25        & 96.3    \\
T1 axial               & 14        & 450              & 3            & 3:25        & 96.1    \\
FIESTA (flip)          & 14        & 4.5              &              & 2:07        & 108.5   \\
Inject Gd&&&&&\\
T1 axial - fat         & 14        & 550              & 3            & 4:09        & 97.1    \\
T1 cor - fat           & 14        & 550              & 3            & 4:09        & 97.4    \\
\hline
Brain target\\
\hline
T1 axial               & 17        & 500              & 5            & 2:12        & 97.3    \\
T2 coronal             & 10.4      & 525              & 5            & 1:43        & 98.1    \\
\end{tabular}
\caption{The complete auditory nerve assessment protocol normally performed by Bloomsburg Hospital.  The $\rm{L_{eq}}$ noise levels displayed in the last column are not A-weighted and do not include application of the flexible tubing acoustical transfer function.  }
\end{table}

Comparisons can also be made between our measured noise levels and those from previous MRI acoustical noise studies.  For example, Randomskij and coworkers measured a very similar noise dosage (per OSHA) of 91.3 dBA from another 1.5 Tesla GE Signa high speed scanner\cite{Randomskij}.  Using otoacoustic emissions, they also discovered that patients experienced significant temporary threshold shifts after MRI examination (with or without hearing protection).  It was further concluded that patients with very sensitive hearing could develop permanent threshold shift as a result of exposure to this MRI noise.  In another study\cite{Counter}, measurements were made of a Bruker Biospec 47/40 experimental MRI system in which the acoustic noise reached a similar peak of 125 dBA. This team also concluded that exposure to repetitive MRI noise in excess of 100 dB would cause auditory threshold shifts.

Auditory threshold shifts are accompanied by several distinct and sometimes rapid physiological changes that occur within the inner ear during exposure to intense noise levels.  Some of these physiological changes are well documented in the literature: damage to the inner hair cells (IHCs), outer hair cells (OHCs) and the stereocilia\cite{Ballenger}.  Other physiological effects are more subtle and still under active investigation: vacuolization and swelling of afferent auditory nerve dendrites at the IHC connections, swelling of spiral ganglion cells and/or their satellite cell sheaths in Rosenthal's canal, and hypertrophe of the stria vascularis\cite{Wang}.

For example, in a recent histological study\cite{Wang}, Wang and coworkers exposed CBA CaJ mice to octave-band noise (8-16 kHz) for a two-hour duration for noise levels between 94 and 116 dB.  (CBA CaJ \cite{jacksonlab} mice have low genetic variability from mouse to mouse to minimize experimental error.) Acute swelling of the inner ear stria vascularis was documented 24 hours after the exposure to 116 dB.  The size of this swelling was on the scale of 50 $\mu$m.  Also observed were hypertrophic swelling of spiral ganglion cells and IHCs that peaked one day and 0.1 days after exposure and at noise levels of 100 and 116 dB SPL, respectively.  It is important to stress that this study was for continuous, steady noise.  By comparison, the MRI noise we measured was temporally varying as indicated by our measured exceedance levels (11 dBA difference between $\rm{L_{10}}$ and $\rm{L_{90}}$).

The noise employed in the murine study by Wang was similar in level to the noise produced by the 1.5 Tesla GE Signa MRI used in our study.  For example, the FIESTA sequence used to target the internal auditory canal was found to produce repetitive maximum noise levels of 108.5 dBA $\rm{L_{eq}}$.  Therefore, it is likely that inner ear physiological changes occur during a FIESTA scan.  Temporarily inflamed tissues are often accompanied by a local increase in water and a decrease in macromolecule concentration\cite{Newhouse}.  This is the same signature used by FIESTA in the detection of acoustic neuromas.  However, as previously stated, ILSs are extremely small and determination of an ILS using MRI could potentially be overlooked and perhaps attributed to a temporary inflammatory change (false negative)\cite{Green}.  We note that the reverse could also occur:  the hypertrophic changes within the cochlea associated with intense acoustical assault could interfere with the diagnosis of a hyperplastic ILS or ICS to the extent that a false positive diagnosis could result.  This is significant in light of the rapid progress in imaging sub-millimeter ILSs and ICSs.  We note that the noise-induced 50 $\mu$m swelling of the murine stria measured by Wang et al. would scale to a 150 $\mu$m swelling of the human stria. The current standard prescription following positive schwannoma diagnosis is one of ``watchful waiting," especially in the case of older individuals for whom neuroma growth rates are slower.  Given the issues we raise, this would seem to be an especially prudent course of action.  False positive diagnoses, although currently rare (however, see footnote\cite{falsepositive}), could become more common if routine schwannoma screening became the norm, as has been proposed by some\cite{Fukui}.

Our work is corroborative of prior work on MRI acoustical noise which focused mainly on the potential for temporary or permanent human hearing threshold shifts\cite{Counter,Randomskij,Ulmer,McJury}.  Beyond this, our study suggests that noise-induced temporary hypertrophe of portions of the intralabyrinthine and intracochlear auditory nerve and inner ear is possible during and after an MRI scan of the brain/VCN.  Because MRI resolution is limited in this region and as the composition of ILSs, ICSs and inflamed tissue is similar, a misdiagnosis cannot be ruled out. Care should therefore be taken during schwannoma assessment (especially within the cochlea) to ensure that any observed masses are in fact persistent and/or growing abnormal hyperplasia (i.e. true schwannomas).

We gratefully acknowledge the generous assistance of Ms. Wendy Panas of Bloomsburg Hospital as well as helpful discussions with Dr. Jorge Gonzalez and Dr. Qing Yue of the Department of Audiology and Speech Pathology at Bloomsburg University of Pennsylvania.  This work was partially supported by the College of Science and Technology of Bloomsburg University of Pennsylvania.


\begin{thebibliography}{99}
\bibitem{falsepositive} In a medical case a 53 year old woman was complaining of hearing aches, disequilibrium, and nausea.  A CAT scan of the IAC was performed and appeared normal\cite{Donnelly}.  An MRI of the IAC taken with a Gd-DTPA enhancement revealed a four mm enhancing lesion at the distal end of the left internal auditory meatus which was assumed to be an acoustic schwannoma.  Surgery was performed, with a 7mm drill out of the IAC to expose the lateral end.  After a meticulous search no tumor was found.  The only abnormality was inflamed arachnoid matter for a distance of 4-5 mm on the vestibular nerve.  This supports the idea that not all MRI scans are correct in diagnosing acoustic neuromas. False negatives as well as false positives can occur\cite{House}.
\end{thebibliography}
\end{document}